# Implementing Pseudofractal Designs in Graphene-Based Quantum Hall Arrays using Minkowski–Bouligand Algorithms

Dominick S. Scaletta,[1] Ngoc Thanh Mai Tran,[2,3] Marta Musso,[4] Dean G. Jarrett,[2] Heather M. Hill,[2] Massimo Ortolano,[4] David B. Newell,[2] and Albert F. Rigosi[2,a]

[1]*Department of Physics, Mount San Jacinto College, Menifee, California 92584, USA*

[2]*Physical Measurement Laboratory, National Institute of Standards and Technology (NIST), Gaithersburg, Maryland 20899, USA*

[3]*Joint Quantum Institute, University of Maryland, College Park, Maryland 20742, USA*

[4]*Department of Electronics and Telecommunications, Politecnico di Torino, Torino 10129, Italy*

[a] Author to whom correspondence should be addressed.  email: afr1@nist.gov

This work introduces a pseudofractal analysis for optimizing high-resistance graphene-based quantized Hall array resistance standards (QHARS). The development of resistance standard device designs through star-mesh transformations is detailed, aimed at minimizing element count. Building on a recent mathematical framework, the approach presented herein refines QHARS device concepts by considering designs incorporating pseudofractals (which may be expressed as star-mesh transformations). To understand how future QHARS pseudofractal designs enable varying sizes of neighborhoods of available quantized resistance, Minkowski–Bouligand algorithms are used to analyze fractal dimensions of the device design topologies. Three distinct partial recursion cases are explored in addition to the original full recursion design, and expressions for their total element counts are derived. These partial recursions, assessed through their fractal dimensions, offer enhanced flexibility in achieving specific resistance values within a desired neighborhood compared to full recursion methods, albeit with an increased number of required elements. The formalisms presented are material-independent, making them broadly applicable to other quantum Hall systems and artifact standards.

# I. INTRODUCTION

Epitaxially grown graphene (EG) has been used to fabricate devices for electrical metrology because of its exhibition of the quantum Hall effect at less demanding experimental conditions with respect to gallium-arsenide devices, partly because of its separation from the SiC beneath via a buffer layer [1-4], and its specific use as a resistance standard entails access to the $v = 2$ Landau level ($R_\text{H} = \frac{1}{2}\frac{h}{e^2} \approx 12906.4037$ Ω). This single-value constraint limits the general measurement infrastructure with which one may define the unit of the ohm. Two main approaches in the literature sought to remove these limitations: (1) quantum Hall array resistance standards (QHARS), interconnecting multiple Hall elements, and (2) the use of *p-n* junctions, with both avenues yielding resistances of $qR_\text{H}$ (with $q$ a positive rational number) [5-17].

Currently, producing large, high-quality epitaxial graphene, essential for creating numerous Hall bar elements in any QHARS devices, is limited to the centimeter scale [18]. For instance, such scales may only enable a several-hundred-element series array, capable of outputting about MΩ level resistances. These values are significantly less than the range up to PΩ-level resistances calibrated globally [19-21]. Future QHARS devices might overcome this limitation by employing star-mesh transformations and pseudofractal designs to achieve much higher resistances [22-27].

This work expands on a recently established framework seeking to minimize the required number of elements in a QHARS device to achieve high effective quantized resistances [22]. The range of resistances for this work will go beyond what was experimentally established (with data from an exemplary QHARS device presented to support the underlying principles of the framework). What is not generally understood in these QHARS device design endeavors is the benefit of using partial recursion to modify the rate at which one arrives at or navigates a neighborhood of desired values (namely, by introducing minor device modifications like adding or removing single elements in a way that is not symmetrically applied to the whole design).

Understanding the correlation between the complexity of a design and its likelihood of flexibility to design changes on part of the experimenter may prevent scenarios in which a device design is rendered less applicable to metrological purposes. The original case of full recursion and three cases of partial recursion (all four of which are pseudofractal designs) will be analyzed in the context of their fractal dimensions (also known as Minkowski–Bouligand dimensions), with the hope of identifying advantages to future device designs. Given that these formulations are independent of a resistance standard's material properties, they may thus be applied to other material systems that exhibit the quantum Hall effect, as well as artifact standard resistors.

## II. METHODS

The QHARS devices used to support this framework (valued near 1 GΩ) were fabricated from square SiC chips measuring 7.6 mm × 7.6 mm that underwent a silicon sublimation procedure described in Ref. [28]. The four main steps that summarize device development are growth, fabrication, post-fabrication, and packaging. EG was grown in a furnace and inspected using optical and confocal laser scanning microscopy [29], followed by fabrication of superconducting device contacts composed of NbTiN [18, 30]. QHARS devices were then measured in a cryostat at approximately 2 K with a Dual Source Bridge (DSB) [23, 26].

The following mathematical analysis further explores the framework for utilizing star-mesh QHARS device designs by using partial recursions for the sake of correlating how such systems behave with fractal dimension and to show practical design considerations that allow for more flexibility in accessing custom quantum resistance values. One needs to recall some of the fundamental principles and conclusions of Ref. [22]. The mathematical relationship between a star network with $N$ terminal nodes and resistances $R_\alpha$ between each node and the star center and its equivalent mesh network with resistances $R_{ij}$ between nodes $i$ and $j$ ($N$ is equal in both networks, but the mesh contains one fewer node) is:

$$R_{ij} = R_i R_j \sum_{\alpha=i}^{N} \frac{1}{R_\alpha}$$

(1)

In Eq. 1, the indices go as high as $N$ with the condition that $i \neq j$. To simplify how a QHARS device undergoes minimal-element design optimization, let us define $q \equiv \frac{R}{R_H}$, where $q$ is defined as the number of single Hall elements held at EG's $\nu = 2$ quantum Hall plateau to obtain the total resistance $R$. Note that this coefficient $q$, the *coefficient of effective resistance* (CER), is typically restricted to the set of positive integers ($q: q \in \mathbb{Z}^+$). One may rewrite Eq. (1) as the following expression:

$$q_{ij} = q_i q_j \sum_{\alpha=i}^{N} \frac{1}{q_\alpha}$$

(2)

And building on the framework from Ref. [22], including all definitions presented therein, the parameter $M$, or recursion number, was applied to the subscripts in such a way that the intended number of cleanroom-fabricated elements is represented by $q_{M:i}$ (single index) and the *effective* (approximate or desired, depending on situation) number of elements is represented by $q_{M:ij}^{(approx)}$ (two indices). Recall, with $\xi$ being the number of grounded branches [22]:

$$q_{M:i} = \frac{1}{\xi}\left(\xi q_{M:ij}^{(approx)} + 1\right)^{2^{-M}} - \frac{1}{\xi}$$

(3)

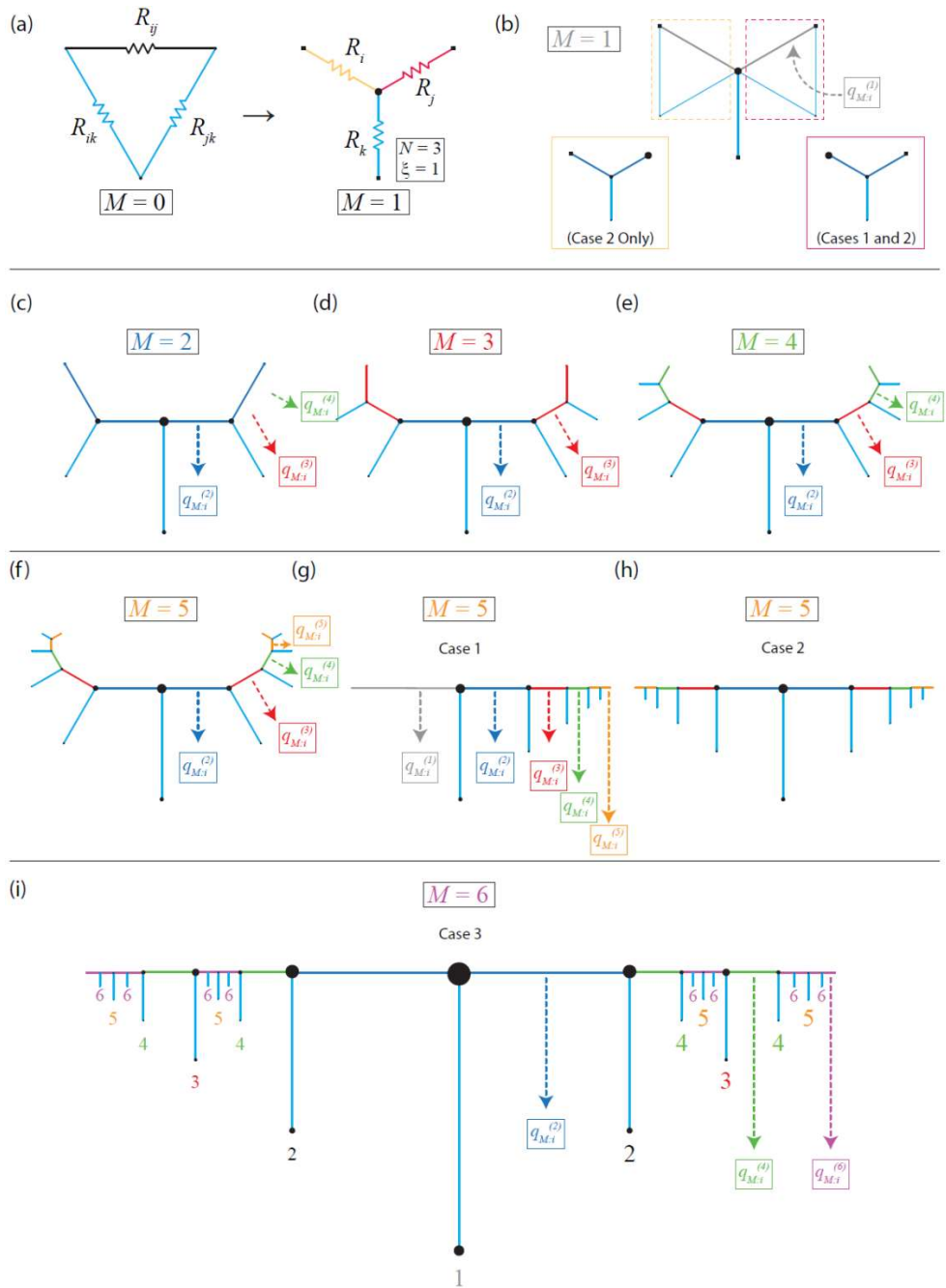

FIG. 1. (a) $R_{ij}$ (equivalently, $q_{ij}$) as a Y-Δ network. Every subsequent expansion of some subset of non-grounded elements increases the characteristic recursion factor $M$ by one. (b) Every resistor in the non-grounded path is expanded as a Y-Δ network, with the full substitution shown in (c). The recursions (for Case 2) are applied for (d) $M = 3$, (e) $M = 4$, (f) $M = 5$, and are done so to grasp the symmetric nature of the recursion. (g) Case 1 ($M = 5$) and (h) Case 2 ($M = 5$) are shown in alternate, topologically similar diagrams used in Ref. [22] for simplicity. (i) Case 3 is drawn up to $M = 6$, where only even numbered recursions are applied fully from the previous (odd) recursion. For odd numbered recursions, Case 2 is replicated.

And that the total number of elements in the final QHARS device is:

$$D_T(M, \xi, q_{M:ij}) = \frac{2^M}{\xi}\left(\xi q_{M:ij}^{(approx)} + 1\right)^{2^{-M}} - \frac{2^M}{\xi} + (2^M - 1)\xi$$

(4)

The pseudofractals that may be generated via this fully recursive treatment from Ref. [22] will be labeled Case 0 since it has been lightly investigated. Three additional pseudofractals reflecting partial recursion that are being investigated are labeled Cases 1, 2, and 3 and are all illustrated in Fig. 1. The iterations for Case 2 are applied for $M = 3$, 4, and 5 to grasp the symmetric nature of the recursion. Topological diagrams are adopted in Fig. 1 (g) and (h) for simplicity [22], and Case 3 is drawn up to $M = 6$, where only even numbered recursions are applied fully from the previous (odd) recursion, and for odd numbered recursions, Case 2 is replicated.

## III. RESULTS AND DISCUSSION

To grant a level of validity to the notion that pseudofractal designs are applicable to quantum electrical metrology, exemplary QHARS devices were fabricated using the design shown in Fig. 2 (a) and (b) (accompanied by a topological drawing used for later pseudofractal analysis). The analog electronic circuit simulator LTspice was used to calculate the exact resistance expected: 1.095 069 ... GΩ [31 – 33]. A precision measurement of the device was then performed with a dual source bridge (DSB), enabling measurement accuracies on the order of μΩ/Ω. The DSB has historically been used as one primary method for measuring high resistances and is essentially an adapted Wheatstone bridge [34-36]. The configuration involves determining an unknown resistance value $R_x$ via $R_x = R_s \frac{V_x}{V_s}$, where $R_s$ is a standard resistor and $V_x$ and $V_s$ are the applied voltages across $R_x$ and $R_s$, respectively. In this case, a standard resistor ($R_s$) valued at approximately 1 GΩ was calibrated with a conventional quantized Hall resistance through a traditional traceability chain [34, 37 – 39]. This standard resistor was then used determine the value of the QHARS device, with the deviation of the output of the device from its predicted value of 1.095 GΩ being called $\delta R_1^{ref}$. These DSB data are shown in Fig. 2 (c) and further support the utility of pseudofractal designs.

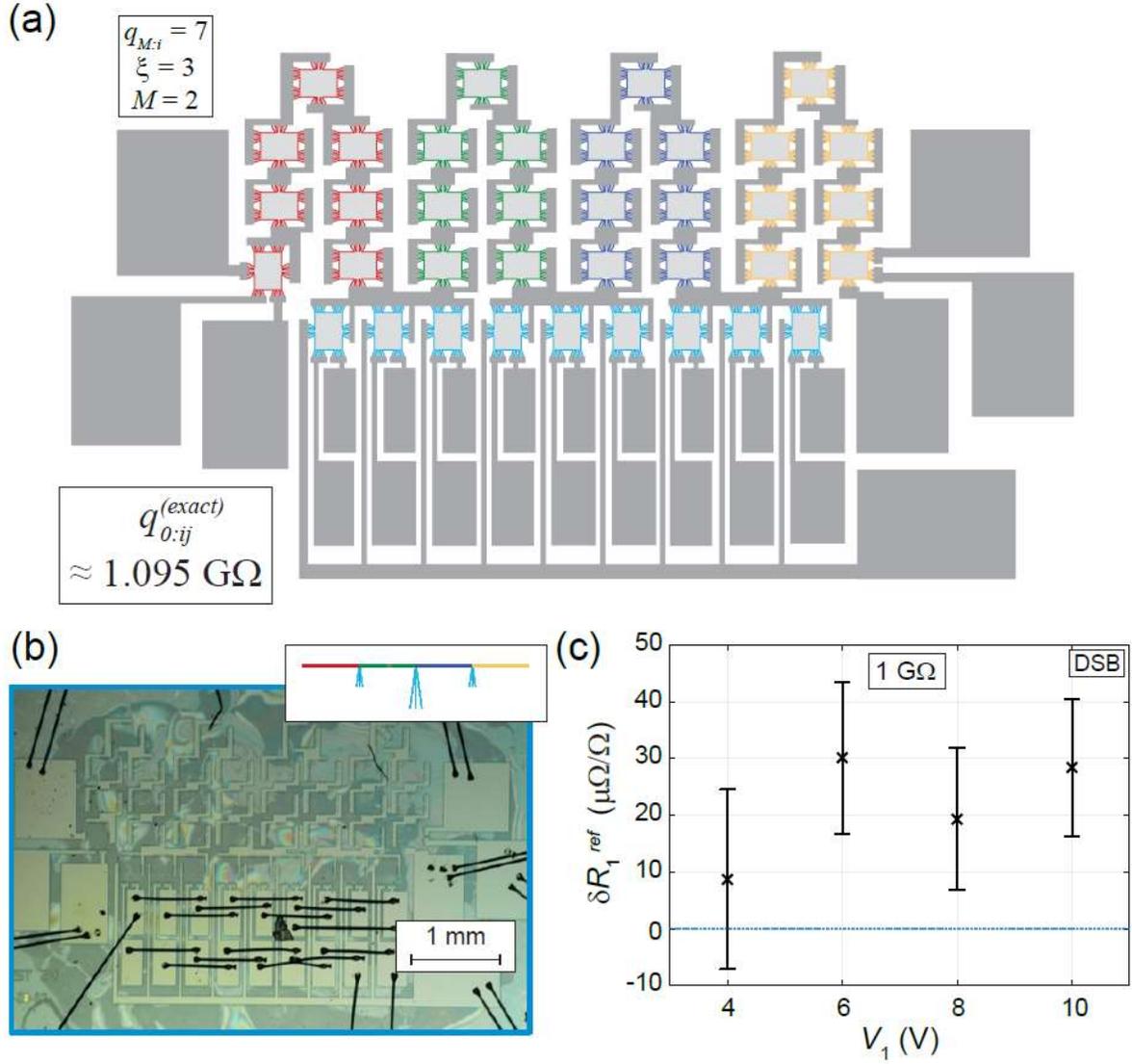

FIG. 2. (a) The layout of the QHARS device design is drawn with corresponding parameters $M = 2$, $\xi = 3$, and $q_{M:i} = 7$. (b) An optical image shows the device post-fabrication and after packaging. An upper inset shows the topological drawing of the QHARS design. (c) Precision measurement DSB data are shown to confirm the quantization of the QHARS device, with the deviation of the device's output from the predicted value of the device, $\delta R_1^{ref}$, centered around $q_{0:ij}^{(exact)} = 1.095\,069\ldots$ G$\Omega$. All error bars in this paper represent a 1σ uncertainty.

Given that pseudofractal designs may provide benefits to metrology, it then becomes important to understand how a pseudofractal's complexity contributes to the feasibility of designing a device with particular features, be it fewer Hall elements or clustered resistance neighborhoods accessible by minor design modification. To this effect, one can modify the definition of $M$ to account for different pseudofractal cases. For Case 1, where $M$ indicates the recursive expansion of only one of the two existing branches most previously dealt with, one introduces the notation $q_{M:i}^{(x)}$, with a superscript $(x)$ to indicate the actual number of elements for the relevant level of recursion (see Fig. 1). This notation does not change how one calculates the number of elements for any specific topological segment:

$$q_{M:i}^{(x)} = \frac{1}{\xi}\left(\xi q_{M:ij}^{(approx)} + 1\right)^{2^{-x}} - \frac{1}{\xi}, x = \{1, 2, \ldots, M\}$$

(5)

However, the total number of elements in a QHARS device changes substantially:

$$D_T(M, \xi, q_{M:ij}) = M\xi + \frac{1}{\xi}\left(\xi q_{M:ij}^{(approx)} + 1\right)^{2^{-M}} - \frac{1}{\xi} + \sum_{x=1}^{M} \frac{1}{\xi}\left(\xi q_{M:ij}^{(approx)} + 1\right)^{2^{-x}} - \frac{1}{\xi}$$

(6)

And this simplifies to:

$$D_T(M, \xi, q_{M:ij}) = M\xi - \frac{(M+1)}{\xi} + \frac{1}{\xi}\left(\xi q_{M:ij}^{(approx)} + 1\right)^{2^{-M}} + \frac{1}{\xi}\sum_{x=1}^{M}\left(\xi q_{M:ij}^{(approx)} + 1\right)^{2^{-x}}$$

(7)

Unlike Eq. 4, which is the main result of Ref. [22], Eq. 7 is discretized as a sum that depends on $M$.

For Case 2, $M$ indicates the recursive expansion of only the outermost existing branches (see Fig. 1). Repeating the analysis yields an identical $q_{M:i}^{(x)}$ as Eq. 5, and the following device count emerges:

$$D_T(M, \xi, q_{M:ij}) = (2M - 1)\xi + \frac{2}{\xi}\left(\xi q_{M:ij}^{(approx)} + 1\right)^{2^{-M}} - \frac{2}{\xi} + 2\sum_{x=2}^{M}\left[\frac{1}{\xi}\left(\xi q_{M:ij}^{(approx)} + 1\right)^{2^{-x}} - \frac{1}{\xi}\right]$$

(8)

For Case 3, $M$ indicates an alternation of two partial recursion patterns as seen in Fig. 1, meant to be a hybrid case of a Cases 0 and 2. Repeating the analysis yields an identical $q_{M:i}^{(x)}$ as Eq. 5, as well as the following device count, which is slightly more unusual due to the alternation:

$$D_T(M, \xi, q_{M:ij}) = 2\xi \sum_{x=1}^{M}\left[2^{\frac{x-H[(-1)^{x+1}]}{2}} - 1\right] + \left(\frac{1}{\xi}\left(\xi q_{M:ij}^{(approx)} + 1\right)^{2^{-M}} - \frac{1}{\xi}\right) * 2^{\frac{M+H[(-1)^{M+1}]}{2} + H[(-1)^M]} + H\left[M - \frac{3}{2}\right]$$
$$* \sum_{x=1}^{M-1} 2^{\frac{x}{2}} * H[(-1)^x] * \left(\frac{1}{\xi}\left(\xi q_{M:ij}^{(approx)} + 1\right)^{2^{-x}} - \frac{1}{\xi}\right)$$

(9)

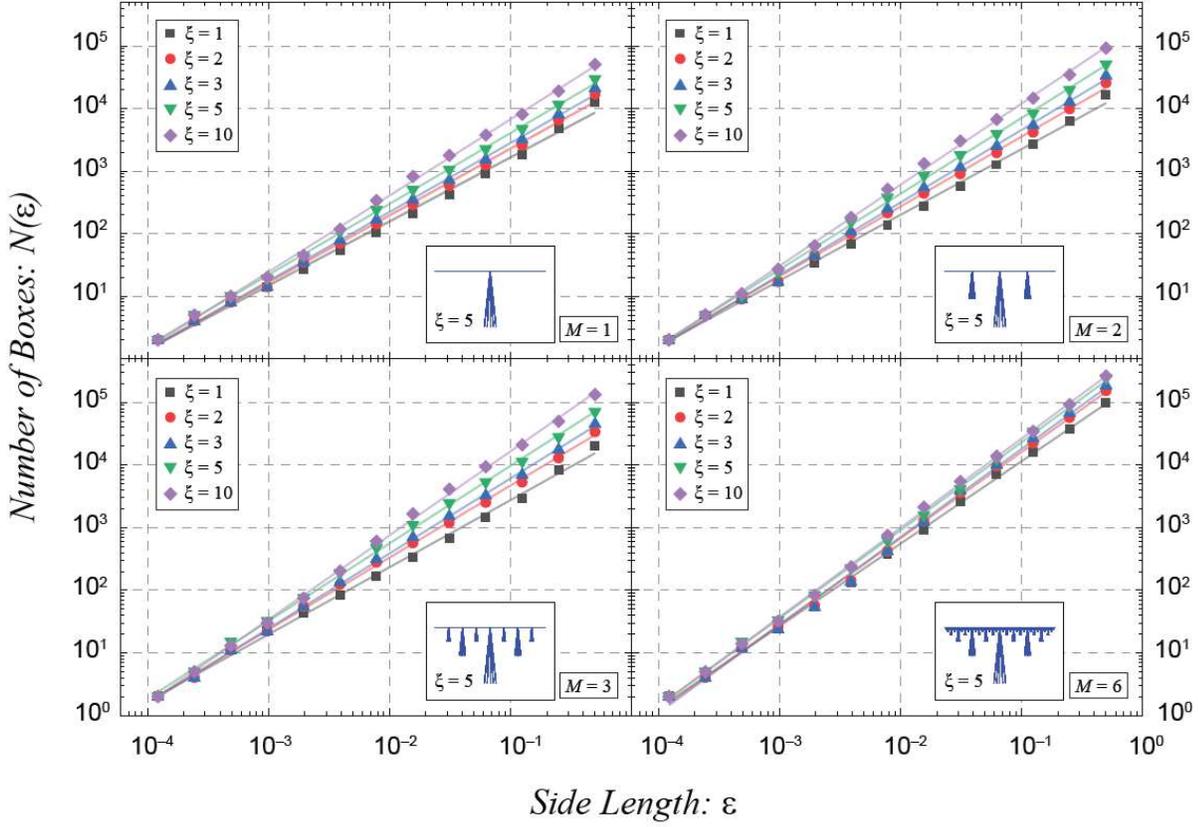

FIG. 3. In Case 0 ($\xi$ = 1, 2, 3, 5, and 10), an example set of calculations are performed via the Minkowski–Bouligand algorithm (also known as the box counting method). An example pseudofractal is shown in each of four lower right-hand insets ($\xi$ = 5), and the four panels reflect several pseudofractal recursions ($M$ = 1, 2, 3, 6). For each set of points, the Minkowski–Bouligand algorithm reveals a fractal dimension upon determining the linear fit on this log–log scale (showing how varying the side length of the boxes used for the algorithm changes the number of boxes containing part of the pseudofractal). Each slope yields a single data point represented in Fig. 4.

In Eq. 9, $H[x]$ is the Heaviside function. Conventionally, the Heaviside function is used to introduce a binary multiplicative operation on various terms to match the alternating nature of the case. For calculations, the function may be substituted by an approximate analytical version with a large corresponding exponent variable $k$: $H[x] \approx \frac{1}{1+e^{-2kx}}$

Now that all four cases have been reviewed, it is important to establish how to quantify the complexity of a pseudofractal QHARS design as represented by its topological drawing in Fig. 3 (and exemplified further in the Supplemental Material [40], along with an explanation of pseudofractal vector quality). Though it is important to note that the specific rendering of the topology of the abstract QHARS design may yield different Minkowski–Bouligand dimensions, the trend on increasing complexity and its correlation with device design flexibility are what should be highlighted. That said, one may define a Minkowski–Bouligand dimension to characterize fractal or pseudofractal set complexity via a ratio of its change in detail to change in scale. For calculating its value, one uses an algorithm to count the number of boxes required for covering the psuedofractal set, assuming it lies on an evenly spaced grid. The number of boxes ($N(\varepsilon)$) changes as the selected grid becomes finer and finer (which gets defined as the box side length $\varepsilon$). From here, the definition of the Minkowski–Bouligand dimension (MBD) arises:

$$dim_{MBD} := \lim_{\varepsilon \to 0} \frac{\log N(\varepsilon)}{\log(1/\varepsilon)}$$

(10)

And from Eq. 10, one can surmise that $N(\varepsilon) \approx C\varepsilon^{-d}$, where $C$ is a constant and $d$ is the MBD. From that point, performing the algorithm yields results as seen in Fig. 3. For the exemplary Case 0 (which includes grounded branch values $\xi$ = 1, 2, 3, 5, and 10), four panels reflect several pseudofractal recursions ($M$ = 1, 2, 3, 6), and an example pseudofractal is shown in each of four lower right-hand insets for $\xi$ = 5. Linear fits are superimposed on the calculations (log–log scale), revealing the MBD for each recursion value ($M$) and number of grounded branches present in the QHARS design topology. Each slope yields a single MBD data point represented in Fig. 4.

Intuitively, the MBD increases in value for nominal increases in complexity, be it from the increased recursions or increase in grounded branches. Figure 4 shows the calculation results for all four Cases as a function of recursion (a similar analysis based on $\xi$ is shown in the Supplemental Material [40]). Each set of data is fitted to a Logistic function defined as such (OriginLab, see Acknowledgments):

$$MBD = A_2 + \frac{A_1 - A_2}{1 + \left(\frac{M}{M_0}\right)^p}$$

(11)

Equation 11, also known as the Hill-Langmuir equation and frequently seen in pharmacology [41], extrapolates the maximum MBD ($A_2$) as $M \to \infty$, and $M_0$ is another fitting parameter indicating the sigmodal center of each fit (see Supplemental Material [40]). The light green and light blue shading indicate the region where the MBD is 95 % and 99.7 % of the maximum value, respectively. With the MBDs fully determined for many QHARS design topologies, one must now correlate this quantity with a metric that suitably characterizes an extent to which a particular topology lends itself to provide a localized or more spread-out neighborhood of resistance values, preferably as a function of the same variables as $M$ or $\xi$.

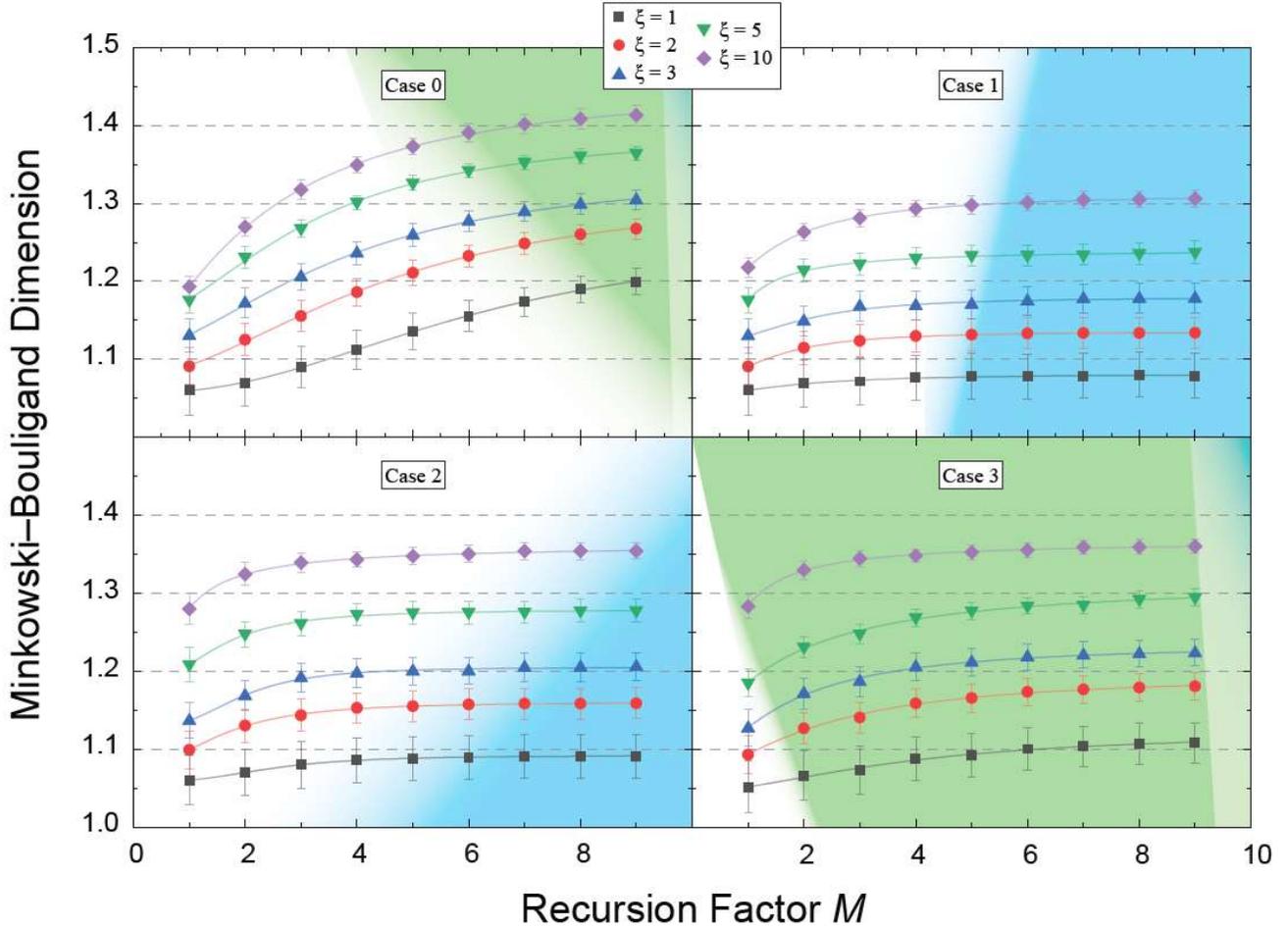

FIG. 4. The Minkowski–Bouligand dimension is determined as a function of $M$ for several $\xi$ and in all four Cases. Each set of data is fitted to a Logistic function as defined in the text to extrapolate the maximum MBD as $M \to \infty$. The light green and light blue shading indicate the region where the MBD is 95 % and 99.7 % of the maximum value, respectively.

One generalized way of devising a design that provides more localized and clustered neighborhoods of available resistances is by carefully inspecting the total device parameter ($D_T$). If this quantity diverges quickly as $M \to \infty$ (i.e. has a high rate of change), then it is safe to conclude that minor modifications (such as the random addition or subtraction of a single Hall element) are more likely to result in larger changes in the output resistance of the QHARS device. Inversely, if the quantity's rate of change is low, then minor modifications are not as likely to result in QHARS output value changes. To better quantify these concepts, each of the four Cases were examined by plotting $D_T$ on a semi-logarithmic scale in Figure 5 (a). Calculations for the adjusted $D_T$ as a function of $M$ reflected pseudofractal designs for resistances including 1 E$\Omega$, 1 P$\Omega$, 1 T$\Omega$, and 10 G$\Omega$ (all values selected to be above what was already experimentally achieved). The top and bottom panels of Fig. 5 (a) focus on $\xi = 1$ and 10, respectively, and all plots are logarithmically normalized such that the minimum values are at 100 for easier comparison. The true total number of devices is 100 added to the corresponding values in the Supplemental Material [40]. In the first panel for each ξ, a dashed orange, partly transparent line is superimposed to extract the exponential coefficient, giving an intuitive grasp on the rate of change of the adjusted $D_T$. Before finally assessing the correlation, it should be noted that, in Fig. 5 (a), Cases 1 and 2 exhibit a rather low rate of change when compared with Cases 0 and 3, regardless of the value of ξ. This can be straightforwardly explained by the limiting behavior of Eqs. 4, 7, 8, and 9 (for Cases 0 through 3, respectively). As $M$ approaches large values in Cases 0 and 3, the behavior is one of exponential growth, but for Cases 1 and 2, it becomes one of linear growth since the terms containing exponential dependences on $M$ approach zero.

With the dependent variable now identified, the Pearson correlation coefficient was then determined between the calculated MBD and the exponential coefficients in (a). The results from this correlation are in Fig. 5 (b) and the coefficient itself is valued at about 0.6379. This positive correlation suggests that for QHARS topologies with higher MBD, one is less likely to design QHARS devices that provide access to localized neighborhoods of resistance. Furthermore, for pseudofractals that are characterized as having linear rather than exponential limiting behavior, QHARS device designers should expect greater ease in providing that localization of values. These two factors should guide designers for future QHARS device topologies.

## IV. CONCLUSIONS

This work provides a solid foundation in the design of QHARS devices by elaborating on the concept of implementing pseudofractal topologies, with the overall concept being supported by experimental data from a fabricated QHARS device. The selection of a pseudofractal as a basis for QHARS device designs has a positively correlated impact on the extent to which a device provides access to a more localized neighborhood of resistance values, as demonstrated by the calculation of the Minkowski–Bouligand dimension for a variety of pseudofractal Cases. Designs that implement pseudofractals with higher MBDs are less likely to maintain a desired output value following minor device design modifications. It then follows that designs exhibiting linear growth in $D_T$ (coupled with a lower MBD) offer better flexibility in accessing desired resistance neighborhoods. These insights into the relationship between design complexity and metrological applicability serve as a guide for future QHARS device development, paving the way for more adaptable resistance standards.

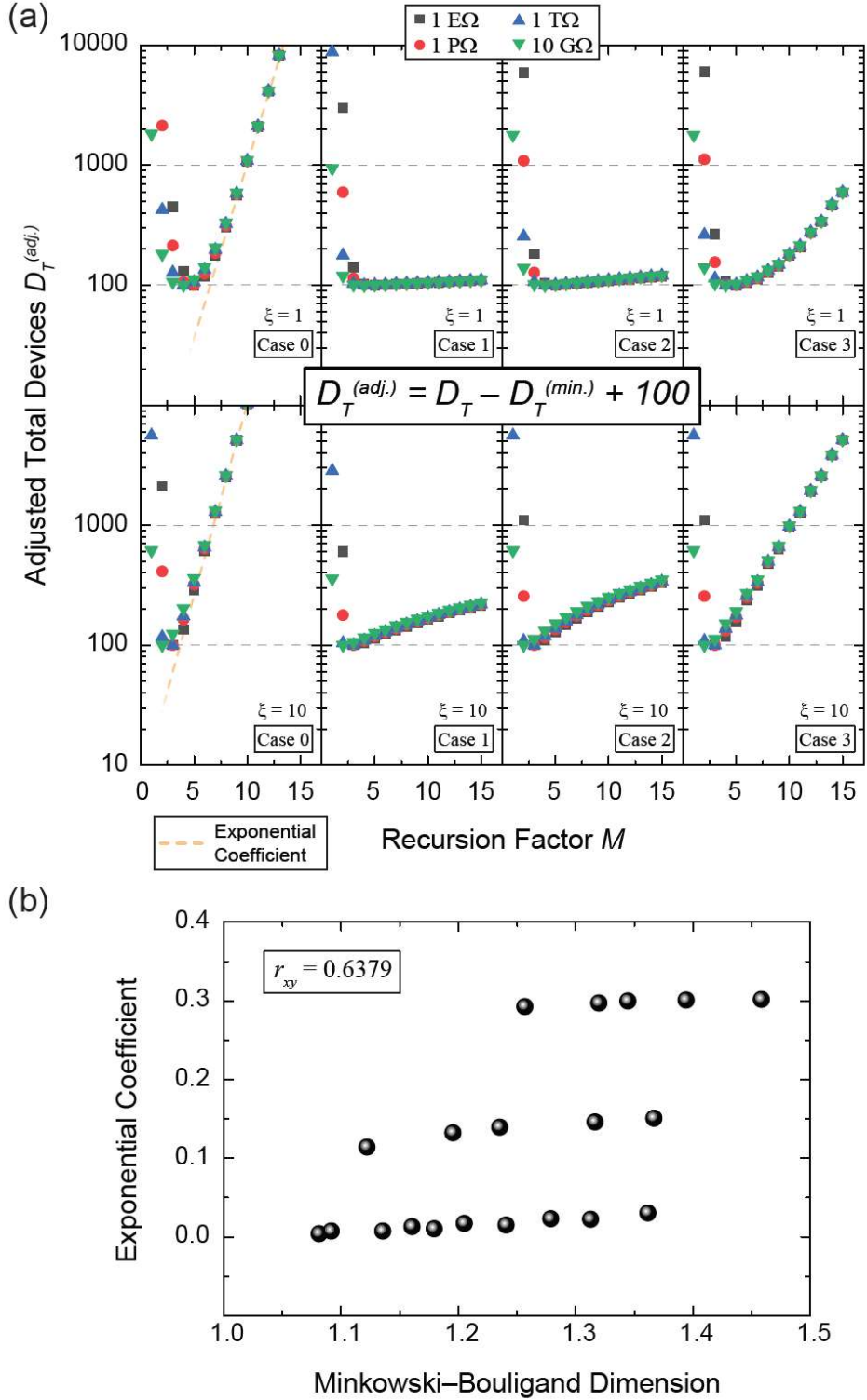

FIG. 5. (a) Calculations for the adjusted total devices as a function of recursion factor ($M$) needed for specific pseudofractal designs. Resistances include 1 E$\Omega$, 1 P$\Omega$, 1 T$\Omega$, and 10 G$\Omega$. The top and bottom panels focus on $\xi = 1$ and 10, respectively, with all four Cases displayed. All plots are logarithmically normalized such that the minimum values are at 100 for easier comparison. The true total number of devices is 100 added to the corresponding values in the Supplemental Material [40]. (b) The Pearson correlation coefficient is determined between the calculated Minkowski–Bouligand Dimension and the exponential coefficients in (a).

## ACKNOWLEDGMENTS

The authors thank M. Munoz, F. Fei, and E. C. Benck for their assistance in the NIST internal review process, and A. R. Panna and Y. Yang for fruitful guidance. The authors declare no competing interests. Commercial equipment, instruments, and materials are identified in this paper in order to specify the experimental procedure adequately. Such identification is not intended to imply recommendation or endorsement by the National Institute of Standards and Technology or the United States government, nor is it intended to imply that the materials or equipment identified are necessarily the best available for the purpose. Work presented herein was performed, for a subset of the authors, as part of their official duties for the United States Government. Funding is hence appropriated by the United States Congress directly. Data that support the findings of this study are available from the corresponding author upon reasonable request.

# Supplemental Material: Implementing Pseudofractal Designs in Graphene-Based Quantum Hall Arrays using Minkowski–Bouligand Algorithms

Dominick S. Scaletta,[1] Ngoc Thanh Mai Tran,[2,3] Marta Musso,[4] Dean G. Jarrett,[2] Heather M. Hill,[2] Massimo Ortolano,[4] David B. Newell,[2] and Albert F. Rigosi[2,a]

[1]Department of Physics, Mount San Jacinto College, Menifee, California 92584, USA

[2]Physical Measurement Laboratory, National Institute of Standards and Technology (NIST), Gaithersburg, Maryland 20899, USA

[3]Joint Quantum Institute, University of Maryland, College Park, Maryland 20742, USA

[4]Department of Electronics and Telecommunications, Politecnico di Torino, Torino 10129, Italy

[b] Author to whom correspondence should be addressed.  email: afr1@nist.gov

Table of Contents:

# 1. Logarithmic Normalization Values for Figure 5

To better compare the behavior of the total device count in the main text, each curve was logarithmically normalized to its global minimum. This table summarizes those values of $D_T^{(min)}$, rounded to the nearest integer:

|  | $D_T^{(min)}$ | | | | | | | |
|---|---|---|---|---|---|---|---|---|
| Grounded Branches (ξ) | 1 | | | | 10 | | | |
| Case | 0 | 1 | 2 | 3 | 0 | 1 | 2 | 3 |
| 1 EΩ | 86 | 8805358 | 6067 | 5984 | 127 | 2.78411E6 | 1134 | 1134 |
| 1 PΩ | 69 | 278913 | 1118 | 1091 | 94 | 88153 | 249 | 249 |
| 1 TΩ | 49 | 8911 | 218 | 212 | 80 | 2832 | 88 | 88 |
| 10 GΩ | 36 | 919 | 79 | 77 | 51 | 309 | 41 | 51 |

## 2. Additional data

Data shown in Figure 5 are complimented with the full set of grounded branch data (Cases 1, 2, and 3, $\xi = \{1, 2, 3, 5, 10\}$):

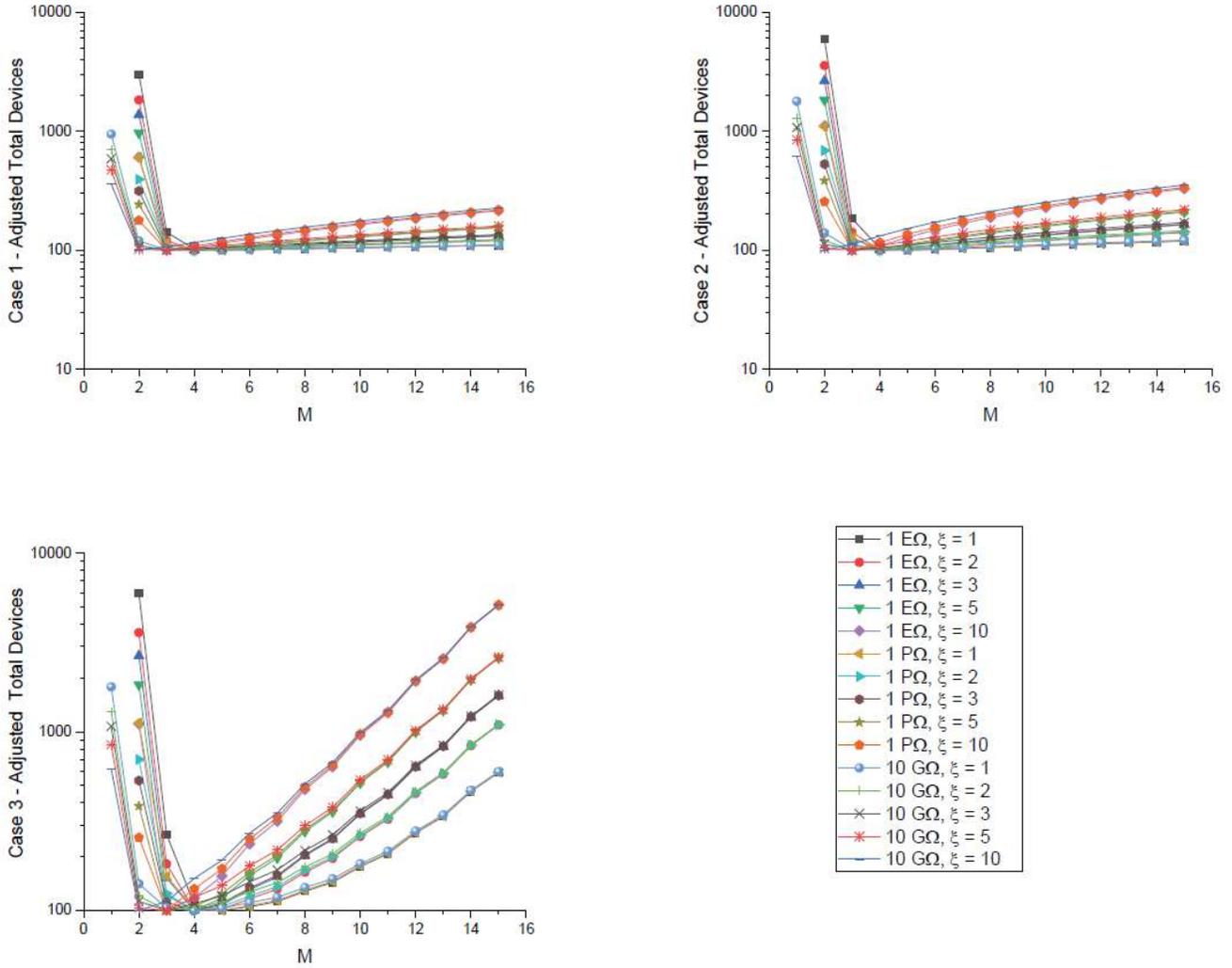

FIG. 1-SM. Calculations for the adjusted total devices as a function of recursion factor ($M$) needed for specific pseudofractal designs. Resistances include 1 E$\Omega$, 1 P$\Omega$, 1 T$\Omega$, and 10 G$\Omega$. There is focus on $\xi = \{1, 2, 3, 5, 10\}$, with three Cases displayed. All plots are logarithmically normalized such that the minimum values are at 100 for easier comparison.

## 3. Additional details about calculating the Minkowski–Bouligand Dimension

To ensure that the vectors of the topological drawings (appended at the end of this document) are not contributing to substantial error when performing the box counting algorithm, a dots-per-inch ([2.54 cm]$^{-1}$) analysis was performed while modifying the quality of the conversion between vector and image. Case 0, $\xi = 10$, was the test scenario and the following trend was found:

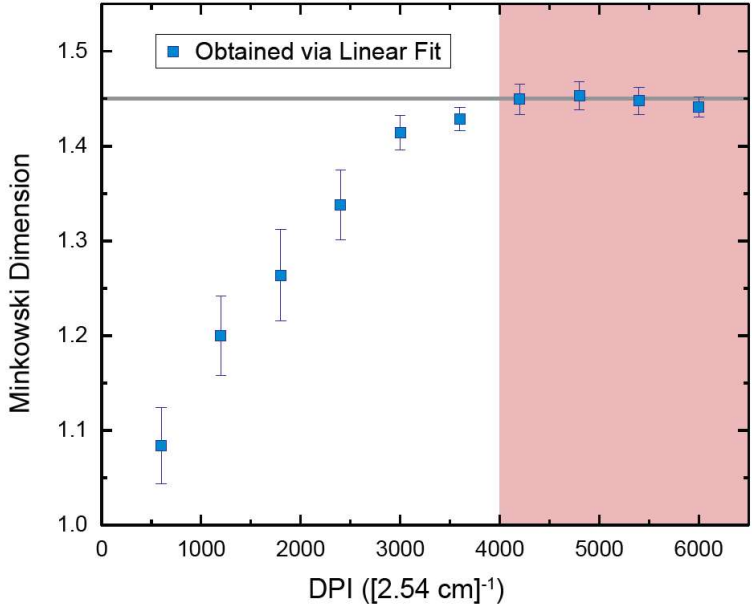

FIG. 2-SM. The DPI test yielded a need to process vectors at least at 4000 DPI (red regime). Doing so ensured that image quality did not contribute to the Minkowski-Bouligand dimension.

And when discussing the MBD, additional information is provided here on the fitting parameter $M_0$, as shown in each fit to a Logistic function defined as such:

$$MBD = A_2 + \frac{A_1 - A_2}{1 + \left(\frac{M}{M_0}\right)^p}$$

(S1)

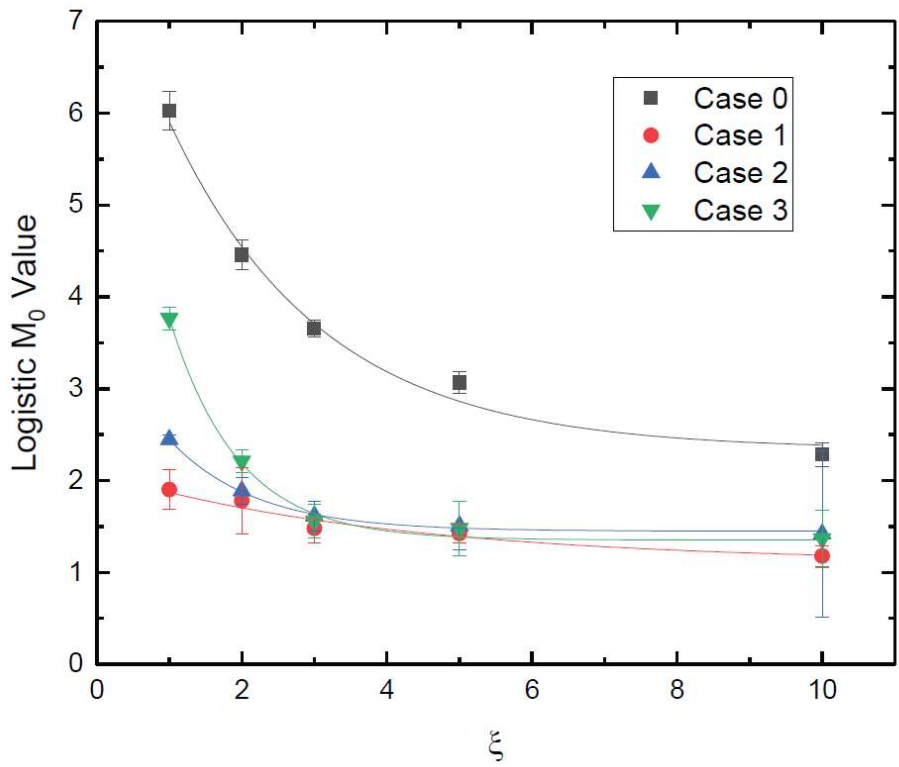

FIG. 3-SM. The sigmoidal center of each Logistic fit (per Eq. S1) generally decay as a function of $\xi$. This can be interpreted as simpler topologies having a greater potential to change their MBD with either $\xi$ or $M$.